\documentclass[aps,pra,showpacs,twocolumn,amsmath,amssymb,superscriptaddress,footinbib]{revtex4}

\usepackage{latexsym}
\usepackage{graphicx}
\usepackage{subfigure}
\usepackage{epsfig}
\usepackage{amsfonts}
\usepackage{amssymb}
\usepackage{amsmath}
\usepackage{bbm}

\usepackage[colorlinks,bookmarks=false,citecolor=blue,linkcolor=red,urlcolor=blue]{hyperref}

\begin{document}

\title{Fractional domain walls from on-site softening in dipolar bosons}

\author{Emma Wikberg}

\affiliation{Department of Physics,
Stockholm University, AlbaNova University Center, 106 91 Stockholm,
Sweden}
\author{Jonas Larson}
\affiliation{Department of Physics,
Stockholm University, AlbaNova University Center, 106 91 Stockholm,
Sweden}
\affiliation{Institut f\"ur Theoretische Physik, Universit\"at zu K\"oln, K\"oln, 50937, Germany}
\author{Emil J. Bergholtz}
\affiliation{Dahlem Center for Complex Quantum Systems and Institut f\"ur Theoretische Physik, Freie Universit\"at Berlin, Arnimallee 14, 14195 Berlin, Germany}
\affiliation{Max-Planck-Institut f\"ur Physik komplexer Systeme, N\"othnitzer Stra\ss e 38, 01187 Dresden, Germany}
\author{Anders Karlhede}
\affiliation{Department of Physics,
Stockholm University, AlbaNova University Center, 106 91 Stockholm,
Sweden}

\date{\today}

\begin{abstract}
We study dipolar bosons in a 1D optical lattice and identify a region in parameter space---strong coupling but relatively weak on-site repulsion---hosting a series of stable charge-density-wave (CDW) states whose low-energy excitations, built from ``fractional domain walls," have remarkable similarities to those of non-Abelian fractional quantum Hall states. Here, a conventional domain wall between translated CDW's may be split by inserting strings of degenerate, but inequivalent, CDW states. Outside these insulating regions, we find numerous supersolids as well as a superfluid regime. The mentioned phases should be accessible experimentally and, in particular, the fractional domain walls can be created in the ground state using single-site addressing, i.e., by locally changing the chemical potential. 

\end{abstract}

\maketitle

\section{Introduction}
The interest in low-dimensional systems, where quantum fluctuations are particularly strong, has increased tremendously during the last decades. Demonstrated by theoretical and experimental studies, systems constrained to one or two spatial dimensions exhibit features related to statistics and topology which are absent in three dimensions. Luttinger liquids, the fractional quantum Hall effect, anyonic and non-Abelian statistics, and graphene are prominent examples. Beyond fundamental interests, these systems might also become instrumental in possible future technological advances such as quantum computing \cite{topo}. 

Adding to the excitement is also the possibility to simulate a rich variety of models using optical lattices, where the dimensionality and interaction parameters are highly controllable. With the newly developed method of trapping cold dipolar atoms or molecules in such lattices~\cite{amexp}, the particles are subject not only to tunneling and contact interaction but also to a long-range interaction. These systems have been studied intensely (see the review~\cite{dipolerev} and references therein), and they display a very rich phase diagram; Mott or charge-density-wave (CDW) states and superfluid and supersolid phases may all be realized by tuning the strength of the on-site interaction, the tunneling amplitude, and the chemical potential.  

In this paper we focus our interest on ultracold bosons (atoms or molecules) in a 1D optical lattice, considering an extended Bose-Hubbard model. The analysis is performed for a long-range repulsive dipolar interaction, as well as for a short-range (on-site plus nearest-neighbor) interaction. Various types of CDW phases have been investigated in the literature earlier; see, e.g., Refs.~\cite{hins,burnell,dima,iskin}. In this work we add  a sequence of exotic members to this list by considering a region in parameter space in which several CDW's of different symmetries become nearly degenerate. While the parameter regions we consider have been investigated in a few earlier (numerical) studies \cite{dima,iskin}, we show that important aspects of the problem have been overlooked, in particular, regarding the entirely new and intriguing excitation structure that emerges in this regime. Using a suggestive analogy to the physics of non-Abelian fractional quantum Hall (FQH) states and an analytical strong-coupling approach assisted by numerical calculations, we find a far more complex and intriguing phase diagram. First, we consider points in parameter space where the problem is exactly solvable, leading to a staircase structure of nontrivially degenerate gapped CDW ground states. Using intuition gathered from recent studies \cite{nonab,ttpfaff} of the Read-Rezayi FQH states \cite{rr}, we find a description of the low-energy excitations in terms of domain walls separating degenerate, translationally inequivalent, CDW's. These domain walls can be created in inequivalent ways and form a representation of $SU(2)_k$ near lattice filling $\nu=k/2$, in close analogy with the excitations of the FQH states \cite{nonab,rr}. Within these insulating regimes, fractional charges (or fractional particle numbers, in cases where the particles are neutral) may be created and manipulated even in the system ground state by locally changing the chemical potential. After having presented this idealized picture, we turn to numerical and perturbative calculations and discuss what happens away from the exactly solvable points, examining different interaction parameters and including tunneling. We also point to some experimental implications.

\begin{figure*}[h!t]
\centerline{\includegraphics[width=1.0\linewidth]{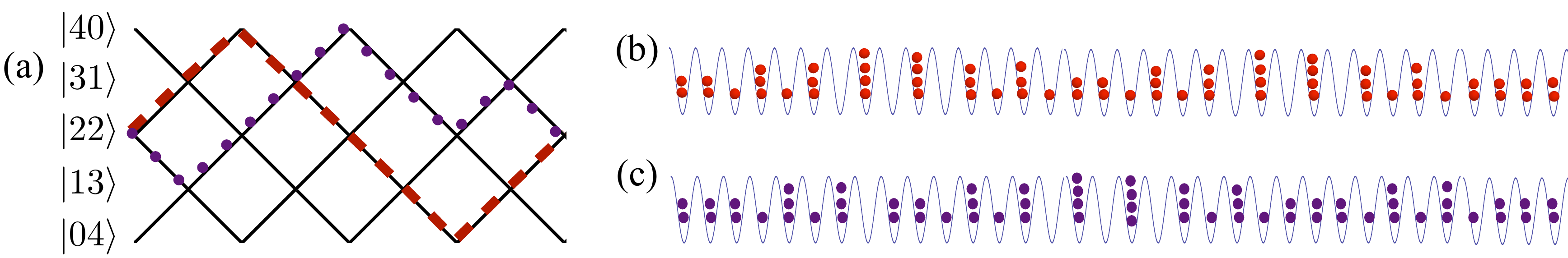}}
\caption{(Color online) (a) How the CDW states, here for $\nu=k/2=2$, can be arranged in a so-called Bratteli diagram. Going from left to right, an elementary domain wall excitation is created in each step. A configuration corresponding to the dashed red line is shown in (b) and a configuration corresponding to the purple dotted line is shown in (c). In (b) and (c) we show only quasi-hole-like domain walls, but quasi-particle-like excitations can be represented analogously.   \label{fig:bratteli}}
\end{figure*}

The outline of the paper is as follows. In the next section we introduce the effective Hamiltonian describing dipolar bosons in 1D. The exact ground state of the system with neither tunneling nor particle-particle interaction beyond nearest neighbors are presented in Sec.~\ref{sec3}. The low-energy excitations are characterized and the analogy of these to the FQH states is outlined. It is, furthermore, shown how our results modify when infinite-range interaction is included. The proceeding Sec.~\ref{sec4} considers the effects arising from weak tunneling between the sites. This analysis is performed using perturbation arguments as well as numerical calculations, and we find that the tunneling in general partially splits the degeneracy between the ground states. Finally we conclude with a discussion in Sec.~\ref{sec5}.

\section{Extended Bose-Hubbard model}\label{sec2}
The system of ultracold bosons in a 1D optical lattice is, after imposing the single-band and tight-binding approximations, described by an extended Bose-Hubbard Hamiltonian;
\begin{equation}
\begin{array}{lll}
\hat{H} & = & \displaystyle{-t\sum_i\left(\hat{a}_i^\dagger\hat{a}_{i+1}+\mathrm{h.c.}\right)-\mu\sum_i\hat{n}_i}\\
& & +\displaystyle{\frac{U}{2}\sum_i\hat{n}_i(\hat{n}_i-1)+\sum_i\sum_{j>0}V_j\hat{n}_i\hat{n}_{i+j}},
\end{array}\label{H}
\end{equation} 
where $t$ is the amplitude for tunneling between neighboring sites, $\mu$ is the chemical potential, $U$ is the on-site interaction, and $V_j$ is the energy cost for a pair of particles $j$ lattice sites apart. If $V_j=V_1/j^3$ is imposed, this accounts for a dipole-dipole interaction between the particles. The bosonic operator $\hat{a}^\dagger_i$ ($\hat{a}_i$) creates (annihilates) a particle on site $i$, while $\hat{n}_i=\hat{a}^\dagger_i\hat{a}_i$ counts the number of particles on site $i$. In this work we neglect any dipole-dipole couplings between neighboring 1D tubes. One of the two transverse directions can be decoupled by properly aligning the dipoles, while coupling in the second transverse direction can be minimized by employing a sharp harmonic trap~\cite{dipolerev}.

As mentioned, the great advantage of utilizing optical lattices is the possibility to tune the many system parameters~\cite{rev}; $t$, $U$, $V_1$, and $\mu$ can all be adjusted using different techniques, like control of the lattice amplitudes, employing Feshbach resonances or alignment of the dipoles. It turns out that the sign of the second derivative of the interaction, $V''_j\equiv V_{j-1}+V_{j+1}-2V_j$ (including $V''_1\equiv U+V_2-2V_1$), is of crucial importance for the low-energy physics of the system~\cite{hubbard,emma}. When all $V''_j>0$, one finds a perfect devil's staircase as a function of the chemical potential---this solution is of relevance both to the fractional quantum Hall effect \cite{hierarchy} and to optical lattices \cite{burnell}. In this paper, we study on-site softening to move away from the above convex case and find a zoo of new phases and exotic domain walls that should be observable experimentally. We will, first, consider a truncated Hamiltonian, keeping only the on-site and the nearest-neighbor interactions, but we will also show that including longer-range terms into the analysis does not change our general findings.

\section{Exact solution}\label{sec3}
Consider first the Hamiltonian in Eq. (\ref{H}) with $t=V_{j>1}=0$. Assuming that the number of particles is allowed to vary, the filling fraction $\nu$ will be fixed by the chemical potential $\mu$. For each such $\mu$, the low-energy physics is completely determined by the relation between $U$ and $V_1$ or, more specifically, by the sign of $V''_1=U-2V_1$. At the point where $V''_1=0$, i.e., where $U=2V_1$, the following holds: For $(k-1)U<\mu<kU$, $k\in\mathbb{N}^+$, the filling fraction is $\nu=k/2$ and all states with periodicity two minimize the energy. In other words, the phase diagram is a staircase where the plateau at $\nu=k/2$ contains the CDW states with unit cells $[l, k-l]$, $l=0,1,...,k$. Note that there is a $(k+1)$-fold degeneracy of the ground state. As an example, when $3U<\mu<4U$, $\nu=2$ and the ground states are the degenerate lattice states $|04\rangle\equiv|0404...\rangle$, $|13\rangle\equiv|1313...\rangle$, $|22\rangle\equiv|2222...\rangle$, $|31\rangle\equiv|3131.
 ..\rangle$, and $|40\rangle\equiv|4040...\rangle$ (here, $\hat n_i|...n_i...\rangle=n_i|...n_i...\rangle$).

\begin{figure}[h]
\centerline{\includegraphics[width=0.99\linewidth]{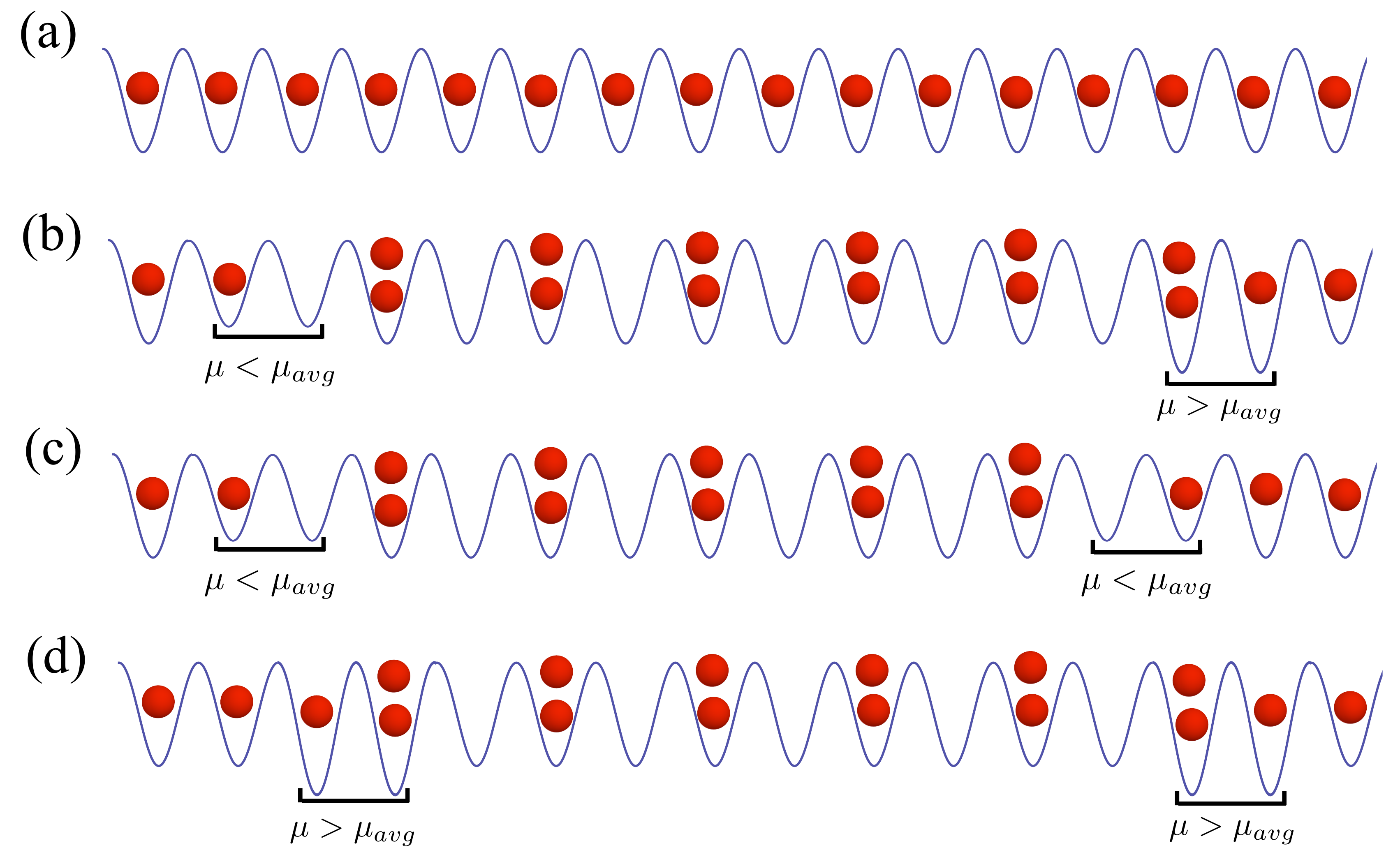}}
\caption{(Color online) Schematic demonstration how a local change in the chemical potential can give rise to fractional domain walls, causing extended changes in the lattice configuration. By increasing or decreasing the local chemical potential on two neighboring sites, one can create domain walls [(b)--(d)] with charge $\pm q/2$, starting from the unperturbed state (a). The domain walls shown in this example are closely related \cite{ttpfaff} to the Majorana fermion excitations of the celebrated Moore-Read FQH state \cite{mr}.} \label{fig3}
\end{figure}

Curiously, precisely these states naturally appear \cite{nonab,ttpfaff,jack,zeros} in the context of the so-called Read-Rezayi \cite{rr} FQH states. Although the Hamiltonian giving rise to the non-Abelian FQH states is starkly different from the one studied here, we find that the lowest-lying excitations of the systems are intimately related. The physical reason for this lies within an emergent ``generalized exclusion principle"; the $\nu=k/2$ Read-Rezayi states can be defined as the unique states within a Landau level that vanish as any $k+1$ particles approach the same coordinate. Analogously, the ground states obtained here are the unique states that never have more than $k$ particles on two consecutive sites. In fact, when the FQH system is put on a thin torus, these two exclusion rules become identical (cf. Ref. \cite{nonab}). Moreover, in both systems the lowest-lying excitations consist of fractionally charged domain walls between (quasi-)degenerate CDW's with unit cells $[l,k-l]$ and $[l\pm1,k-l\mp1]$, respectively, as shown in Figs. \ref{fig:bratteli} and \ref{fig3}  \cite{nonab}. The fractional charge of the excitations is $\pm\frac{q}{2}$, supposing the particles all carry the charge $q$ \cite{com1}; to see this, create two identical domain walls by inserting a string of $|l\pm1,k-l\mp1\rangle$ in a $|l,k-l\rangle$ state, and note that this state has one particle more or less than the ground state \cite{Schrieffer,nonab}. For $k>1$ it is possible to create domain wall excitations with a given charge and position in inequivalent ways, as is demonstrated in the left panel of Fig. \ref{fig:bratteli}, showing a so-called Bratteli diagram \cite{Bratteli}. Here, transitions between different CDW states is illustrated by drawing lines connecting states with $l$ differing by 1. There are several ways to get from the starting to the ending state (the example in the figure being $|22\rangle$), on the way creating a certain number of domain wall excitations. Hence, there is a nontrivial degeneracy of these. A pedagogical account for the precise relation between Bratteli diagrams and topological FQH states can be found in Ref. \cite{nonab}.

The very same nontrivial degeneracy of the excitations is at the heart of the non-Abelian exchange statistics in the corresponding FQH states and provides the crucial non-local degrees of freedom that make these states prime candidates in the quest for topologically protected quantum computation \cite{topo}. However, the analogy between the present setup and the two-dimensional FQH states has the important limitation that we do not have any fractional exchange statistics in our system. Nevertheless, proposals for quantum information processing in 1D quantum wires have been presented~\cite{1dstat}. While we have not worked out some similar idea in the current system, we may note that the present setup has the nice property that more than a single internal electronic level of the atom or molecule can be considered. Coupling between electronic states then allows for transfer of atoms between different atomic lattices, which, in turn, can be spatially separated via adiabatic motion. In this respect, single-site atoms can be ``lifted out'' from the lattice~\cite{bloch2,bloch4}. Utilizing this possibility might be one way to achieve exchange statistics.

In order to have a robust realization of the above introduced fractional domain walls, it is of great interest to know whether these types of states can form in a ground-state configuration as well. The answer is yes. If the experiment is set up so the filling fraction can be forced to deviate slightly from half-filling, the fractional charges will appear in the ground state. If the filling fraction on the other hand is adjusted automatically by the chemical potential and a bath of particles, e.g., in the presence of a confining harmonic trap, the domain walls can instead be created by locally varying the chemical potential. Single-site control of the chemical potential indeed can be achieved via tightly focused lasers, as was recently shown experimentally~\cite{bloch2}. This newly developed technique for locally controlling $\mu$ is most likely superior to earlier proposals relying on impurity atoms~\cite{castin}. We find that increasing $\mu$ slightly on two neighboring sites will tend to create a domain wall with increased particle number, while decreasing $\mu$ on two neighboring sites has the opposite effect. This is illustrated in Fig.~\ref{fig3}.

In a laboratory setting there is inevitably an external trapping potential. This leads to the formation of incompressible domains of different particle densities (each at $\nu=k/2$ for various $k$), separated by relatively sharp boundaries. In each of these incompressible domains the excitation structure follows the discussion above, thus, we expect a given sample to harbor excitations that locally mimic those of different FQH states. 

The formation of domain walls is dependent on the degeneracy of the different CDW's. A relevant question to pose is whether one can tune the interaction so the full degeneracy persists also for the more realistic long-range dipolar interaction $V_j=V_1/j^3$ (we still treat $t=0$). To answer this question, we note that the energy difference between any pair of these states is proportional to $\sum_{j\geq1}V''_{2j-1}$, meaning that if this sum is zero they are all degenerate. Inserting the expression for $V_j$ above, one finds the requirement
\begin{eqnarray}
0&=&\sum_{j\geq1}V''_{2j-1}=U+2V_1\sum_{j\geq1}\frac{(-1)^j}{j^3}\nonumber\\  &=&U-V_1\frac{3}{2}\zeta(3)\approx U-2V_1\times0.9015,
\end{eqnarray} 
where $\zeta(x)$ is the Riemann $\zeta$ function. Hence, the relation between $U$ and $V_1$, which for the truncated Hamiltonian was $U=2V_1$, has only been slightly shifted, and the staircase structure as well as the nature of the fractionally charged domain walls remain qualitatively the same. However, the domain walls now interact weakly due to the long-distance tail of the interaction, and the width and positions of the insulating plateaus are modified. For the dipolar interaction, the plateau corresponding to filling fraction $\nu=k/2$ extends over $U(\frac{7k}{6}-\frac{1}{2}-\frac{\pi^2}{18\zeta(3)})<\mu<U(\frac{7k}{6}-\frac{1}{2}+\frac{\pi^2}{18\zeta(3)})$ or, roughly, $U(1.1667k-0.9561)<\mu<U(1.1667k-0.0439)$. Between these plateaus, which notably do not cover the entire range in $\mu$, we argue that the fractional domain walls described above form a devil's staircase of their own---these minimally charged excitations repel each other and form ordered ground states at any rational filling fraction, just like the underlying particles do for a convex interaction \cite{hubbard,hierarchy,burnell}.

\section{Away from the exactly solvable points}\label{sec4}
\subsection{Numerical and perturbative results}\label{sec4a}
The analysis presented in the previous section is exact for $t=0$ and when the relative strengths of $U$ and $V_1$ are such that the CDW states are exactly degenerate. What happens as we relax one of these requirements by altering the relationship between $U$ and $V_1$? We note that the domain walls do not interact at the degeneracy point $U=2V_1$. Slightly away from this point, one finds that they interact via a weak linear potential, whose strength and sign depends on the energy splittings between the CDW's. However, the nature of the domain walls and their fractional charge do not change.

\begin{figure}[h]
\centerline{\includegraphics[width=0.99\linewidth]{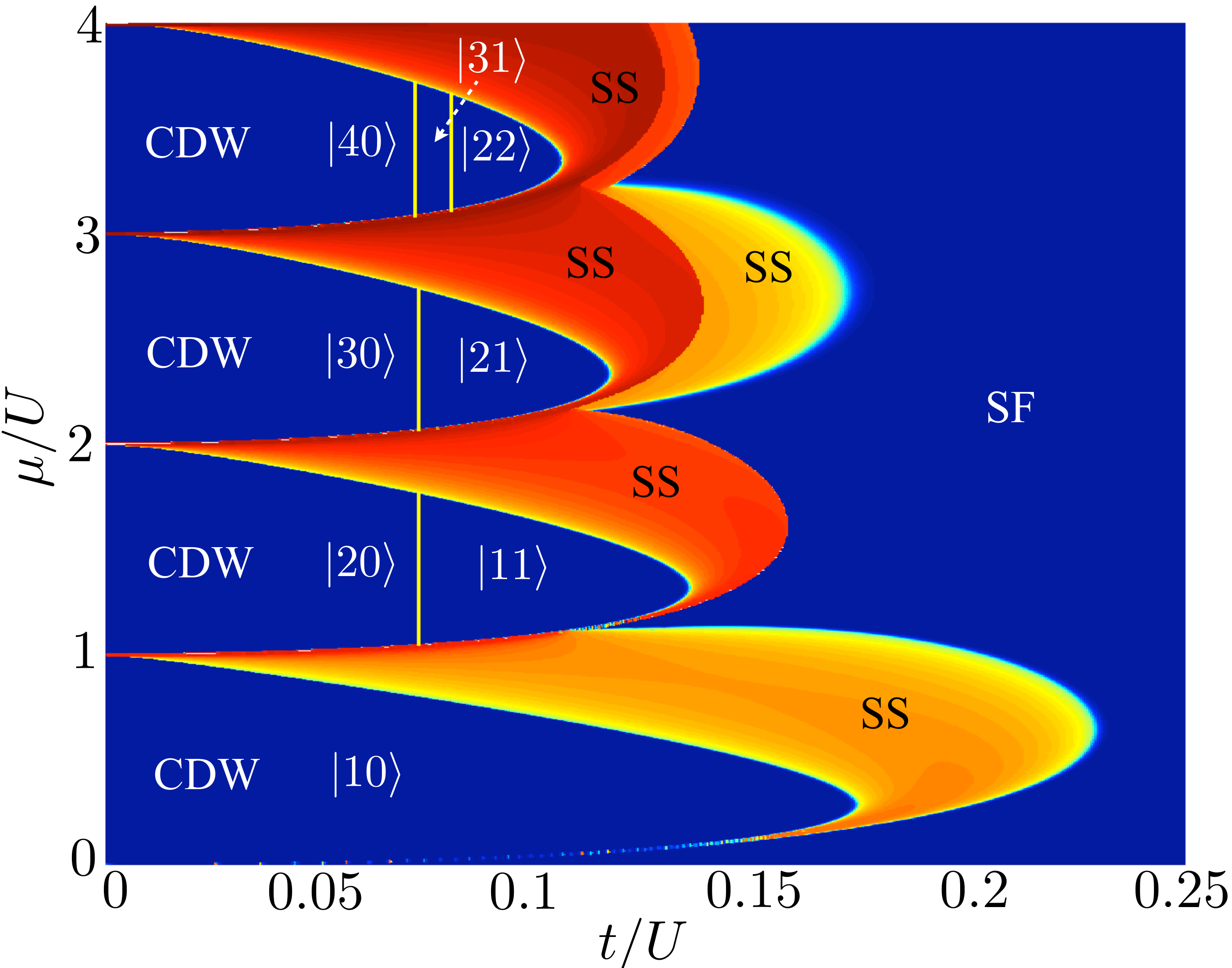}}
\caption{(Color online.) Phase diagram for $U=1.99V_1$, $V_{j>1}=0$, obtained by employing a Gutzwiller ansatz, together with perturbation theory up to third order in $t/U$. The yellow vertical lines represent the phase transitions between the different CDW states, and the diagram also displays the transitions between the insulating CDW regions, the supersolid (SS) phases and the superfluid (SF) region. The colorscale marks only the different phases.} \label{fig1}
\end{figure}

In actual experiments, there is always some finite possibility for tunneling between different sites in the lattice. In this section we present analytical and numerical calculations indicating what happens as we include the nearest-neighbor tunneling term ($t>0$) in the Hamiltonian. Our results are summarized in the phase diagram in Fig.~\ref{fig1}, which is based on a numerical analysis using the Gutzwiller-ansatz method~\cite{gutzwiller}, together with a perturbative strong-coupling expansion up to third order in $t/U$. Even though it is known that the Gutzwiller mean-field approach is not giving a quantitatively correct ground state in 1D~\cite{sce}, we believe that, together with the perturbative results, it indeed gives a qualitatively accurate picture of the phase diagram. 

The degeneracy of the CDW states with unit cells $[l,k-l]$ will, in general, be split by the introduction of the tunneling parameter $t$. In a strong-coupling expansion~\cite{sce}, we include the energy corrections obtained using perturbation theory up to third order, and discarding proportionality factors and redundant constants, the energies of the CDW's as a function of $l$ become
\begin{equation}
\begin{array}{lll}
E_l^0+E_l^2 \sim (l^2-kl)(U-2V_1)+\\
 +2t^2\Bigg{\{}\!\frac{l(k-l+1)}{U(2l-k-1)+V_1(2k-4l+1)}\!+\!\frac{(k-l)(l+1)}{U(k-2l-1)+V_1(4l-2k+1)}\!\Bigg{\}},
\end{array}\label{Ecorr}
\end{equation} 
assuming $V_{j>1}=0$. This is correct up to third order in $t/U$ since all odd-order corrections are zero. Due to the form of Eq. (\ref{Ecorr}), it is not possible to make all CDW states exactly degenerate simultaneously. However, for any fixed value of the tunneling parameter $t$, one can always tune the on-site interaction so CDW states with unit cells $[l,k-l]$ and $[l\pm1,k-l\mp1]$ become \emph{pairwise} degenerate, and, thus, the fractionally charged domain wall excitations remain. To be more specific, since all the CDW states are still approximately degenerate for small values of $t$, the low-energy excitations will include all types of domain walls contained in the Bratteli diagram; see Fig. \ref{fig:bratteli}.

Alternatively, keeping the relationship between $U$ and $V_1$ fixed, one can switch between different CDW states by varying $t$, as shown in Fig. \ref{fig1}. Hopping favors states with the particles evenly spread out, which means that one needs to choose $U<2V_1$ in order to see the (first-order) phase transitions between all CDW states; hence Fig. \ref{fig1} is constructed for $U=1.99V_1$. Within the insulating lobes in the phase diagram, the energy difference between any two CDW states at the specific filling fraction is independent of $\mu$. Consequently, the lines separating two adjacent CDW phases are vertical. The positions of the vertical lines in the figure have been calculated using Eq. (\ref{Ecorr}).

The remaining structure of the phase diagram has been obtained numerically using the Gutzwiller-ansatz wave function
\begin{equation}
|\Psi\rangle=\prod_i\left(\sum_{n=0}^Nf_n^{(i)}|n\rangle_i\right),
\end{equation}
where $|n\rangle_i$ is the $n$-atom Fock state at site $i$, $f_n^{(i)}$ is the corresponding amplitude, and $N$ is a truncation number ($N=30$ in our calculations). We restrict the numerics to the most relevant situations consisting of solutions of periodicity two, and, accordingly, the lattice occupations $f_n^{(i)}$ may be separated into two sublattices $A$ and $B$~\cite{iskin}. The values of $f_n^{(A,B)}$ are obtained by minimizing the energy functional $E=\langle\Psi|\hat{H}|\Psi\rangle$ self-consistently. For the given parameters we find a set of different supersolid (SS) phases~\cite{com2} surrounding the insulating CDW's. These SS's are characterized by nonzero  $\phi_\alpha=\langle\hat{n}_\alpha\rangle$ for the two lattices $\alpha=A,\,B$, together with a nonzero order parameter $\Delta\phi=\phi_A-\phi_B$. The different CDW's as well as the SS's are both separated by first-order phase transitions. In the insulator regime it follows that the transitions between CDW crystalline phases have to be of first order since $\Delta\phi$ must be discontinuous at the transition. We believe that the same mechanism drives the transitions between SS phases, and, hence, the transition is, again, of first order. In particular, the SS's can be seen as partially melted CDW phases, and the different SS phases support distinct underlying CDW states. Further away from the insulating regimes, the SS phases turn into pure superfluid phases ($\phi_A=\phi_B\neq0$) via second-order phase transitions. In order to clearly distinguish between the different SS phases, Fig.~\ref{fig1} is constructed based on calculations of the \emph{derivative} $d\Delta\phi/dt$ of the order parameter.

\subsection{Experimental implications}\label{sec4b}
The different regimes in the phase diagram, presented in Fig.~\ref{fig1}, should be accessible experimentally (see discussion in Ref.~\cite{iskin}), opening up the possibility of seeing new types of CDW phases in a 1D lattice. Another prospect discussed in our study concerns the possibility of varying the chemical potential locally to create fractionally charged domain walls. A suggestion might be to switch on the disturbance in $\mu$ adiabatically, using the technique of Ref.~\cite{bloch2}, and then suddenly turn it off again to follow the system evolution as the domain wall delocalizes, giving a measure of its effective mass or the so-called Lieb-Robinson bound~\cite{diss}. Similar measurements were indeed recently performed in a system following a quantum quench~\cite{bloch5}. As for this experiment, the dynamical delocalization can, in principle, be tracked down with the help of individual-site fluorescence methods, recently implemented in various experiments~\cite{bloch3}. One important observation is that the above on-site measurements are limited to detect the ``parity,'' meaning the even or odd number of atoms. Even though there are suggestions regarding how to go beyond this restriction~\cite{bloch_com}, we note that whenever a domain wall separates two different CDW phases, the parity differs between the two phases. Thus, even if the measurement is capable of detecting only the parity, this should already contain information of the propagation of the excitations. An alternative way to create such a nonequilibrium state could be to drive the system nonadiabatically through one of the phase transitions separating the various CDW's and thereby to create defect excitations. 

\section{Concluding remarks}\label{sec5}
We have studied ultracold bosons in a 1D optical lattice, assuming an extended Bose-Hubbard model. For particular choices of the strength of the on-site interaction $U$, we find a nontrivial increased ground-state degeneracy, both for a truncated Hamiltonian and for a long-range dipolar interaction between the particles. In the weak-tunneling regime, the ground states are different CDW states with periodicity two. Within the degenerate regime it is possible to create fractionally charged domain walls between the degenerate vacua by varying the chemical potential locally in the lattice. Including tunneling in the analysis, we presented a phase diagram achieved using perturbation theory together with a numerical simulation assuming a Gutzwiller ansatz, and the type of phase transitions between the different CDW states, as well as between supersolid and superfluid regimes, were identified. We also pointed out that all experimental techniques necessary to verify our results already exist.

In this work we also presented an intriguing relationship between ultracold bosons in 1D and the 2D FQH effect. We showed that on-site softening leads to an emergent ``generalized exclusion principle" that forbids $k+1$ particles to occupy two consecutive sites, in striking similarity to the clustering conditions on the non-Abelian Read Rezayi FQH states that dictates how the wave-function amplitude must vanish as $k+1$ approach the same coordinate. In fact, in both systems this property can be taken as the defining property of the ground states (as well as the quasihole excitations). We also showed that the structure of low-energy excitations, built from ``fractional domain walls," is identical to that of the Read Rezayi states. Moreover, we submit that this ``generalized exclusion principle" should be much easier to directly observe in the 1D dipolar systems discussed here than in the context of non-Abelian FQH states where such concepts are frequently used; see, e.g., Refs. \cite{nonab,ttpfaff,jack,zeros}.

We also want to stress that the analogy between ultracold bosons in 1D and the 2D FQH effect also holds in the case of simpler, Abelian FQH states; the solution of the very same problem giving rise to the devil's staircase for dipolar bosons (without on-site softening) \cite{burnell} was earlier used to provide a microscopic understanding of the self-similar phase diagram observed in the lowest Landau level \cite{hierarchy}.

We hope that this analogy can spur further cross-fertilization between the two fields and we note that it is possible to obtain a whole plethora of nontrivially degenerate CDW ground states by making other choices for various $V''_j$ \cite{emma}.

\begin{acknowledgements}
We thank Maciej Lewenstein and Immanuel Bloch for helpful discussions. J.L. and A.K. acknowledge support from the Swedish research council (VR). E.J.B. is supported by the Alexander von Humboldt foundation. J.L. also acknowledges support from Deutscher Akademischer Austausch Dienst (DAAD) and Kungl. Vetenskapsakademien (KVA). E.J.B. acknowledges useful discussions with Achilleas Lazarides.
\end{acknowledgements}


\begin{thebibliography}{999}

\bibitem{topo}C.~Nayak, S.~H.~Simon, A.~Stern, M.~Freedman, and S.~Das Sarma, 
Rev.\ Mod.\ Phys.\ {\bf 80}, 1083 (2008). 

\bibitem{amexp} A. Griesmaier, J. Werner, S. Hensler, J. Stuhler, and T. Pfau, Phys. Rev. Lett. {\bf 94}, 160401 (2005); K. K. Ni, S. Ospelkaus, M. H. G. de Miranda, A. Peer, B. Neyenhuis, J. J. Zirbel, S. Kotochigova, P. S. Julienne, D. S. Jin, and J. Ye, Science {\bf 322}, 231 (2008); J. G. Danzl, M. J. Mark, E. Haller, M. Gustavsson, R. Hart, J. Aldegunde, J. M. Hutson, and H. C. N\"agerl, HC , Nature Phys. {\bf 6}, 265 (2010). 

\bibitem{dipolerev} T. Lahaye, C. Minotti, L. Santos, M. Lewenstein, and T. Pfau, Rep. Prog. Phys. {\bf 72}, 126401 (2009).

\bibitem{iskin} M. Iskin, Phys. Rev. A {\bf 83}, 051606 (2011).

\bibitem{dima} D. L. Kovrizhin, G. Venketeswara Pai, and S. Sinha, Europhys. Lett. {\bf 72} 162 (2005). 

\bibitem{hins}
	E. Berg, E. G. Dalla Torre, T. Giamarchi, and E. Altman,
	Phys. Rev. B {\bf 77}, 245119 (2008). 
	
\bibitem{burnell} F. J. Burnell, M. M. Parish, N. R. Cooper, and S. L. Sondhi, Phys. Rev. B {\bf 80}, 174519 (2009).
	
\bibitem{ttpfaff} E. J. Bergholtz, J. Kailasvuori, E. Wikberg, T. H. Hansson, and A. Karlhede, Phys. Rev. B {\bf 74}, 081308(R) (2006); A. Seidel and D.-H. Lee, Phys. Rev. Lett. {\bf 97}, 056804 (2006).
	
\bibitem{nonab} E. Ardonne, E. J. Bergholtz, J. Kailasvuori, and E. Wikberg, 
J. Stat. Mech. (2008) P04016.

\bibitem{rr}N. Read and E. Rezayi, Phys. Rev. B {\bf 59}, 8084 (1999).

\bibitem{rev} M. Lewenstein, A. Sanpera, V. Ahufinger, B. Damski, A. Sen(De), U. Sen, Adv. Phys. {\bf 56}, 243 (2007).

\bibitem{emma} E. Wikberg (to be published).

\bibitem{hubbard} J. Hubbard, Phys. Rev. B 17, 494 (1978). 

\bibitem{hierarchy} E. J. Bergholtz, T. H. Hansson, M. Hermanns, and A. Karlhede, 
Phys. Rev. Lett. {\bf 99}, 256803 (2007); E. J. Bergholtz and A. Karlhede, Phys. Rev. B {\bf 77}, 155308 (2008).

\bibitem{jack} B. A. Bernevig and F. D. M. Haldane,  Phys. Rev. Lett. {\bf 102}, 066802 (2009).

\bibitem{zeros} M. Barkeshli and  X.-G. Wen, Phys. Rev. B {\bf 79}, 195132 (2009).

\bibitem{com1} In case the particles are neutral, this translates into fractional particle number.

\bibitem{Schrieffer} W. P. Su and J. R. Schrieffer, Phys. Rev. Lett. {\bf 46}, 738 (1981).

\bibitem{Bratteli} O. Bratteli, Trans. Amer. Math. Soc. {\bf 171}, 195 (1972).

\bibitem{1dstat} J. Alicea, Y. Oreg, G. Refael, F. von Oppen, and M. P. A. Fisher, Nature Physics {\bf 7}, 412 (2011).

\bibitem{bloch2} C. Weitenberg, M. Endres, J. F. Sherson, M. Cheneau, P. Schauss, T. Fukuhara, I. Bloch, and S. Kuhr, Nature {\bf 471}, 319 (2011).

\bibitem{bloch4} O. Mandel, M. Greiner, A. Widera, T. Rom, T. W. H\"ansch, and I. Bloch, Phys. Rev. Lett. {\bf 91}, 010407 (2003).

\bibitem{mr} G. Moore and N. Read, Nucl. Phys. B {\bf 360}, 362 (1991).

\bibitem{castin} U. Gavish and Y. Castin, Phys. Rev. Lett. {\bf 95}, 020401 (2005).

\bibitem{gutzwiller} J. Zakrzewski, Phys. Rev. A {\bf 71}, 043601 (2005).

\bibitem{sce} J. K. Freericks and H. Monien, Europhys. Lett. {\bf 26}, 545 (1994).

\bibitem{com2} Off-diagonal long-range order is, in its strict sense, forbidden in 1D systems [N. D. Mermin and H. Wagner, Phys. Rev. Lett. {\bf 17}, 1133 (1966)], but an algebraically decaying long-range order is possible and, therefore, we use the term supersolid. 

\bibitem{diss} J. Dziarmaga, W. H. Zurek, and M Zwolak, Nature Phys. {\bf 8}, 49 (2012).

\bibitem{bloch3} W. S. Bakr, J. I. Gillen, A. Peng, S. F\"olling, and M. Greiner, Nature {\bf 462}, 74 (2009); J. F. Sherson, C. Weitenberg, M. Endres, M. Cheneau, I. Bloch, and S. Kuhr, Nature {\bf 467}, 68 (2010).

\bibitem{bloch5} M. Cheneau, P. Barmettler, D. Poletti, M. Endres, P. Schauß, T. Fukuhara, C. Gross, I. Bloch, C. Kollath, and S. Kuhr, Nature {\bf 481}, 484 (2012).

\bibitem{bloch_com} I. Bloch (private communications).



\end{thebibliography}
\end{document}